\documentclass{aa}  

\usepackage{graphicx}
\usepackage{txfonts}
\usepackage{url}
\usepackage{CJK}

\begin{document} 

\begin{CJK*}{UTF8}{bsmi}

   \title{Anatomy of the internal bow shocks in the
   IRAS 04166+2706 protostellar jet}


   \titlerunning{Internal bow shocks in the IRAS 04166+2706 jet}

   \author{M. Tafalla\inst{1}
   \and Y.-N. Su\inst{2} 
   \and H. Shang\inst{2} 
   \and D. Johnstone\inst{3,4}
   \and Q. Zhang\inst{5} 
   \and J. Santiago-Garc\'{\i}a\inst{6}
   \and C.-F. Lee\inst{2}
   \and N. Hirano\inst{2} 
   \and L.-Y. Wang (王亮堯) \inst{2,7}
   }

   \institute{Observatorio Astron\'omico Nacional (IGN), Alfonso XII 3, 
              E-28014 Madrid, Spain 
              \email{m.tafalla@oan.es}
         \and
  Academia Sinica, Institute of Astrophysics (ASIAA), and Theoretical
  Institute for Advanced Research in Astrophysics (TIARA), Academia Sinica,
  11F of Astronomy-Mathematics Building, AS/NTU. No.1, Sec. 4, Roosevelt Rd,
  Taipei 10617, Taiwan, R.O.C.
         \and
National Research Council of Canada, Herzberg Astronomy \& Astrophysics, 
5071 West Saanich Road, Victoria, BC, V9E 2E7, Canada 
         \and
Department of Physics and Astronomy, University of Victoria, Victoria, BC V8P 1A1, Canada
         \and
Harvard-Smithsonian Center for Astrophysics, 60 Garden Street, Cambridge, MA 02138, USA
         \and
Instituto de Radioastronom\'{\i}a Milim\'etrica (IRAM), Avenida Divina
Pastora 7, N\'ucleo Central, 18012 Granada, Spain
         \and
Graduate Institute of Astronomy and Astrophysics, National Taiwan
University, No. 1, Sec. 4, Roosevelt Road, Taipei 10617, Taiwan, R.O.C.         }

             \authorrunning{M. Tafalla et al.}

   \date{Received ; accepted }

 
  \abstract
   {Highly collimated jets and wide-angle outflows are two related 
   components of the mass-ejection activity
   associated with stellar birth. Despite decades of research, the relation
   between these two components remains poorly understood.}
   {We study the relation between the jet and the outflow in the
   IRAS~04166+2706 protostar. This Taurus protostar drives a molecular jet 
   that contains multiple emission peaks symmetrically located from the central source. 
   The protostar also drives a wide-angle outflow consisting of two conical shells.}
   {We have used the Atacama Large Millimeter/submillimeter Array (ALMA) 
   interferometer to observe two fields along the
   IRAS~04166+2706 jet. The fields were centered on a pair of  
   emission peaks that 
   correspond to the same ejection event. The observations were carried out in 
   CO(2--1), SiO(5--4), and SO(J$_{\mathrm N}$=6$_5$--5$_4$).}
   {Both ALMA fields present spatial distributions that are approximately elliptical and have
   their minor axes aligned with the jet direction.
   As the velocity increases, the emission in each field moves
   gradually across the elliptical region.
   This systematic pattern indicates that the emitting gas in each field
   lies in a disk-like structure that is perpendicular to the jet axis and whose
   gas is expanding away from the jet. A small degree of curvature in the first-moment 
   maps indicates that the disks are slightly curved in the manner expected for bow
   shocks moving away from the IRAS source. A simple geometrical model confirms
   that this scenario fits the main emission features.}
   {The emission peaks in the IRAS~04166+2706 jet likely represent internal bow shocks
   where material is being ejected laterally away from
   the jet axis. While the linear momentum of the ejected gas is dominated by the
   component in the jet direction, the sideways component is not negligible,
   and can potentially affect the distribution of gas in the surrounding outflow 
   and core.}


   \maketitle
%

\section{Introduction}

Highly collimated jets and wide-angle molecular outflows are common signatures
of stellar birth. They are thought to represent two aspects of the same mass-loss 
phenomenon and to arise from the need of a protostar to release angular momentum
as it contracts.
Despite decades of study, however, the exact relation between these two forms of 
ejection is still a matter of debate \citep{fra14,arc07}.
One popular view supports the view that the jets are the denser, inner parts of
wide-angle protostellar winds, and that these wider winds are the true accelerating
agents of the large-scale outflows \citep{shu00,sha06}.
A popular alternative defends the view that the
jets are the sole driving agents of the outflows, and that there is
no need to invoke an unseen wide-angle component to explain the less-collimated 
appearance of the molecular flows. According to this view,
the action of the narrow jets can be widened by 
several mechanisms, including jet precession or wandering,
entrainment,
and the lateral ejection of material in internal jet shocks \citep{mas93,sta94,rag93}.

While observations of individual outflows often support either the
wide-angle wind or jet-only picture,
systematic studies of outflow morphology and kinematics have shown that no single
model can explain the variety of observations \citep{lee00}.
Compounding this problem is the rarity of systems where both the highly 
collimated jet and the molecular outflow can be studied simultaneously. 
This limitation partly results from
an observational bias, since the detection of a jet, because of its atomic
composition, needs to be carried out at optical or infrared wavelengths,
which 
favors objects with little obscuration. However, the detection of a molecular
outflow is achieved in the radio, and therefore
requires a highly embedded system \citep{che95}.

Fortunately, 
observations have slowly revealed that there is a small group of 
protostars where both the
jet and molecular outflow can be studied simultaneously. 
These protostars, probably due to their extreme
youth, have highly collimated jets that are
molecular instead of atomic, and therefore can be observed
with the same molecular tracers as those used to study
the wider angle outflows.
Examples of these young protostars are L1448-mm and IRAS 04166+2706
(IRAS~04166 hereafter). These systems
present molecular spectra with characteristic secondary
components of extremely high velocity (EHV) gas that are
distinctly separated from the standard outflow wings
\citep{bac90,taf04,taf10}. When the EHV
components are observed with high angular resolution,
they are found to trace 
highly collimated molecular jets that travel
inside conical cavities whose walls represent the
lower velocity, wide-angle outflows 
\citep{gui92,bac95,san09,hir10,wan14}.
Although L1448-mm and IRAS 04166 represent the finest examples of
molecular jets, they are not the only examples. Systems like
HH~211 and HH~212 also present highly collimated molecular jets,
but their low inclination angle with respect to the plane
of the sky hinders the detection of clearly detached EHV features in
the spectra \citep{gue99,pal06,lee07,cod14,lee15}.

In this paper we present Atacama Large Millimeter/submillimeter Array (ALMA) observations of 
the molecular jet driven by IRAS~04166. 
This protostar is a 0.4~L$_\sun$ class 0 source 
in the Taurus molecular cloud at an estimated distance of 140~pc \citep{eli78,tor07}.
The outflow of this protostar presents a remarkable jet-plus-cavity morphology. 
Interferometric observations 
by \citet{san09} and \citet{wan14} (SG09 and W14 hereafter) 
have resolved the IRAS~04166 jet into a
collection of discrete emission peaks that are located symmetrically with 
respect to the IRAS position, and which likely correspond to a number of 
pulsation events in the history of the jet ejection.
Position-velocity diagrams along the jet axis made by these authors
reveal that each 
emission peak has an internal linear velocity gradient where
the gas that appears closer to the source moves faster than the 
gas that appears further away.
This type of velocity gradient matches the prediction from models
of pulsating jets (e.g., \citealt{wil84,rag90,sto93}), 
which show that variability in the velocity of ejection at the
jet base creates a series of internal shocks where
rapidly moving jet material
overtakes slower gas launched previously.
When these shocks occur, the gas inside the jet is 
ejected sideways and the projection of this motion along the line
of sight produces a characteristic
saw-toothed pattern (e.g., Fig.~16 in \citealt{sto93}).

Unfortunately, the observations of the IRAS~04166 jet by SG09 and W14 lacked 
the necessary sensitivity to study the 2D velocity field of the
gas in the EHV emission peaks, and the interpretation of the lateral ejection motions
had to rely on the analysis of 1D position-velocity diagrams.
The recent online availability of ALMA, with its significant increase in sensitivity
and image quality,
has now made it possible to study the 2D velocity structure of
the relatively weak jet emission in detail.
In this paper, we present ALMA observations of the IRAS~04166 jet 
aimed to determine the full velocity field of two selected EHV emission peaks. 
By using the ALMA interferometer, and by concentrating on
only two jet positions, these observations were designed to achieve 
high enough sensitivity to study the internal motions of the EHV gas.
As shown below, the new observations not only
confirm the previous interpretation of the kinematics of the EHV gas,
but provide a detailed picture of how the gas is laterally ejected 
in this remarkable molecular jet.

\section{Observations}

\begin{figure}
\centering
\resizebox{\hsize}{!}{\includegraphics{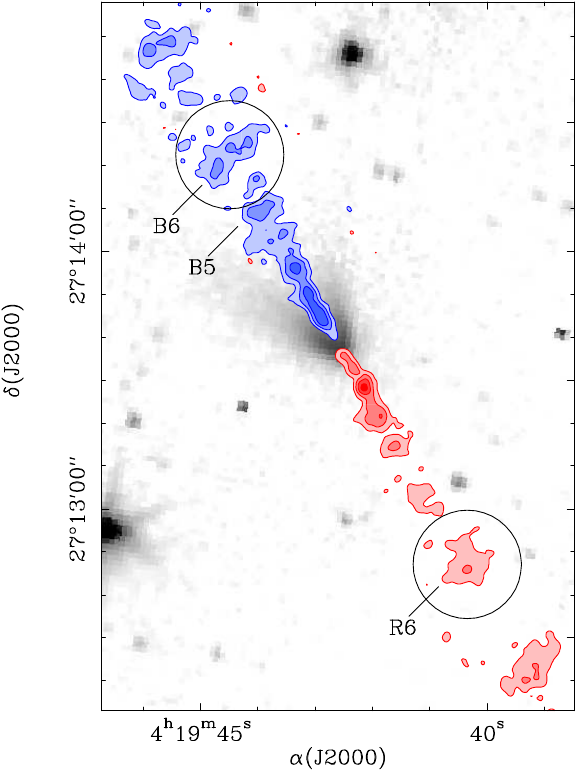}}
\caption{Large-scale image of the IRAS 04166 jet 
with circles representing the ALMA target positions.
The color scale and contours represent the
CO(2--1) EHV emission observed by SG09 with the IRAM PdBI, and the 
background grayscale is an archival Spitzer/IRAC1 image. 
The two $25''$diameter circles represent the fields of view of the ALMA
antennas. The B5, B6, and R6 labels indicate the emission peaks identified
by SG09 and further studied in this paper.}
\label{fig1}
\end{figure}

We used ALMA to observe the two target fields indicated with circles in Fig.~1.
These fields are
located at offsets ($26\farcs5$, $46\farcs5$) and 
($-28\farcs8$, $-48\farcs9$) with respect to the position of 
IRAS 04166 at $\alpha(J2000)=4^h19^m42\fs5,$ $\delta(J2000)=+27^\circ13'36''$
(SG09). They were observed jointly during two 1.3-hour ALMA scheduling blocks
in December 2014, as part of
Early Science Cycle 1 project 2012.1.00304.S (PI Yu-Nung Su). 
The Band 6 receivers 
were tuned to a frequency of 230.5~GHz, to observe simultaneously CO(J=2--1),
SiO(J=5--4), and SO(J$_{\mathrm N}$=6$_5$--5$_4$). The correlator was 
configured to provide a velocity resolution of approximately 0.32~km~s$^{-1}$.
The telescope array was in its C32-2 configuration and the baselines between its 
12 m antennas ranged approximately between 15 and 330~m. Because of shadowing limitations
caused by the low declination of the source, no observations with the compact Morita Array were made for this project.

The ALMA observations consisted of alternating pointings on the two target fields
interspersed with observations of J0510+1800 for phase calibration. The same
continuum source was used for flux calibration, and 
J0423-0120 was used to calibrate the bandpass.
The resulting visibility data were reduced using the observatory pipeline in
CASA version 4.3.1 \citep{mac07}, which was used for calibration, image synthesis, and
preliminary cleaning. Imaging was carried out using briggs weighting with a 
robust parameter of 0.5, and resulted in a synthesized beam of
$1\farcs5 \times 1\farcs1$ for the higher frequency CO(2--1) line (230.5 GHz) and 
$1\farcs6 \times 1\farcs2$ for the lowest frequency SiO(5--4) line (217.1 GHz).
Final cleaning of the maps was carried out using the MAPPING
program of the GILDAS software\footnote{\url{http://www.iram.fr/IRAMFR/GILDAS}},
although no significant differences were found between the cleaning results 
from CASA and MAPPING. 
The typical rms noise level in the maps is 2.5~mJ~beam$^{-1}$ (= 35~mK) 
per 0.5~km~s$^{-1}$ channel.

\section{Results}

Fig.~\ref{fig1} shows, with circles, the location of the two ALMA fields on a colored image of 
the EHV CO(2--1) emission obtained with the IRAM PdBI.
The fields were chosen to
cover a pair of emission peaks labeled B6 and R6 in the SG09 notation,
and these seem to have resulted 
from an ejection event that occurred about 900~yr ago. They were
selected as ALMA targets because they are well separated from the other emission peaks
within the jet, and because their size matches the primary beam of the 12 m ALMA antennas
($25''$ FWHM).
In this paper, we refer to these as the ``northern'' and ``southern'' fields,
and unless indicated, we color code the first one in blue and the second one
in red following the color of the outflow lobe to which they belong.

\begin{figure}
\centering
\resizebox{\hsize}{!}{\includegraphics{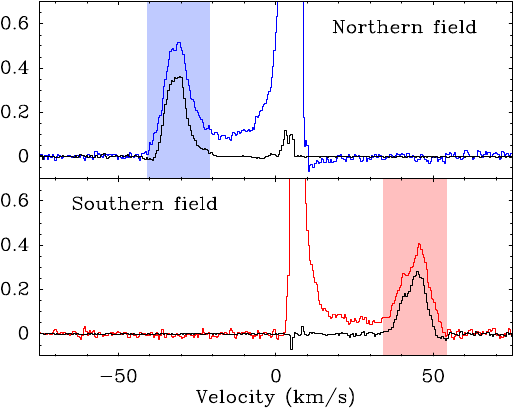}}
\caption{Comparison between the IRAM 30 m single dish spectra
from T04 (blue and red histograms) and ALMA spectra synthesized
for the same positions
and angular resolution ($11''$). The color vertical bands indicate the
location of the blue and red EHV regimes.
There is a good agreement between the 30 m and ALMA spectra
inside the EHV ranges, which is indicative of only minor missing flux.
IRAM 30 m data in $T_\mathrm{mb}$ scale and ALMA data in $T_\mathrm{B}$ scale.}
\label{fig2}
\end{figure}

The ALMA observations did not include the compact array, and as a result, they
may be missing flux from the largest spatial scales.
For this reason, we start our analysis by
comparing the ALMA observations with single-dish data, which contain all the 
flux and thus provide a reliable reference.
We use the CO(2--1) line because it is the brightest and
most extended line in the ALMA observations and, therefore, the most sensitive 
line to the missing flux problem.
As for single dish data, we use the spectra from \citet{taf04} (T04 hereafter),
who mapped the CO(2--1) emission of the IRAS 04166 outflow
with the IRAM 30 m single-dish telescope. From these data we chose the
two positions closest to each of the ALMA fields, and
we compared these positions with
spectra synthesized from the ALMA data assuming the same position and
angular resolution as the 30 m observations ($11''$ FWHM).
Fig.~\ref{fig2} shows the ALMA-30 m comparison using colored histograms for the 
single-dish data and black histograms for the synthesized ALMA data.

As can be seen, the ALMA spectra match well the single-dish data
for velocities inside the EHV regime (color-shaded 
bands), where the intensity agreement between the two spectra
is within about 30\% at the line peak.
This match implies that the
ALMA observations recover most of the emission inside the EHV regime,
and that the lack of compact array data does not significantly affect the observations of the EHV gas.

\begin{figure}
\centering
\resizebox{\hsize}{!}{\includegraphics{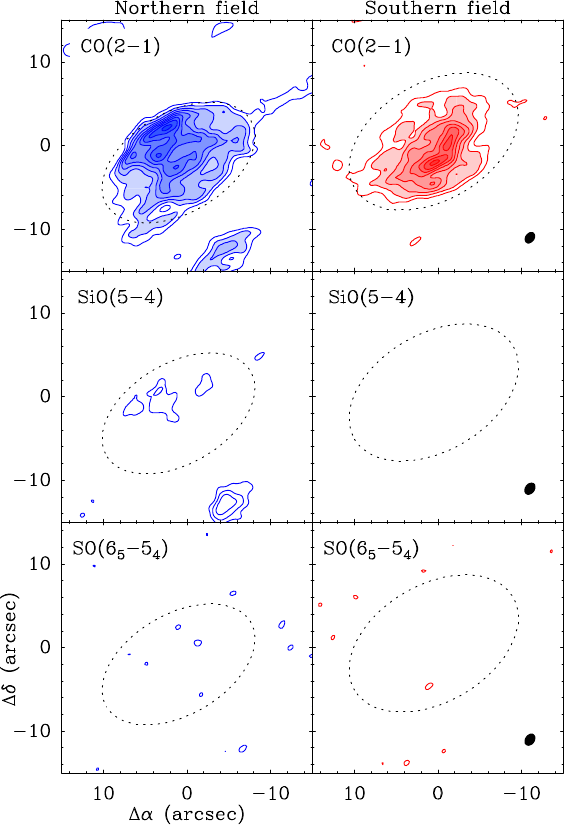}}
\caption{Maps of the integrated EHV emission in the two ALMA target fields for the 
three main 
transitions in the setup: CO(2--1) (upper row of panels), SiO(5--4) (middle row), and
SO(J$_{\mathrm N}$=6$_5$--5$_4$) (bottom row). 
First contour and contour interval are 1.0~K~km~s$^{-1}$ in the CO(2--1) maps
and 0.5~K~km~s$^{-1}$ in the SiO and SO maps.
The dashed ellipses delineate the approximate boundaries of
the CO emission.
The small filled ellipses in the maps of the southern field indicate the synthesized
beam for each transition.
}
\label{fig3}
\end{figure}

In contrast with the EHV regime, the emission from the low-velocity outflow
is mostly missing in the ALMA observations (see velocities near the
systemic value of 6.7~km~s$^{-1}$).
This loss is caused by the very extended nature of this emission, whose size
exceeds the ALMA primary beam, and which can only be
mapped using mosaicking techniques.
For this reason, we do not study the slow outflow component and our analysis is restricted to the EHV regime.

\subsection{Integrated intensity maps}

\begin{figure*}
\centering
\resizebox{\hsize}{!}{\includegraphics{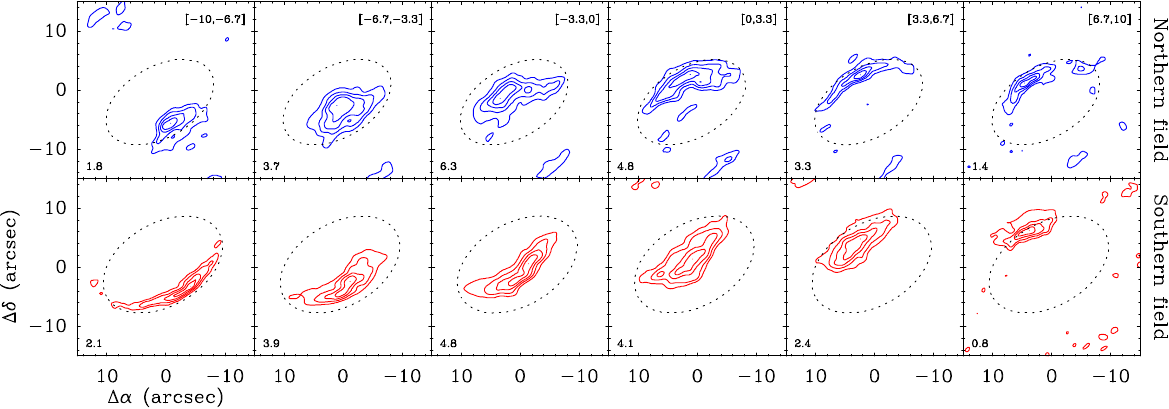}}
\caption{Channel maps of CO(2--1) emission for the northern (top, color-coded blue)
and southern (bottom, color-coded red)
ALMA fields. For each field, the maps cover the 20~km~s$^{-1}$-wide range 
shaded with color in Fig.~\ref{fig2}. The velocity range of each map,
after correction for a jet velocity of 37.5~km~s$^{-1}$, is
indicated in the top row in units of km~s$^{-1}$.
First contour and step are 20\% of the map emission peak, which is indicated for each 
panel in the bottom left corner in units of K~km~s$^{-1}$.
The dashed ellipses delineate the approximate boundary of the emission. The emission from SW to NE is gradually displaced as the velocity increases.
}
\label{fig4}
\end{figure*}

Fig.~\ref{fig3} shows maps of the emission integrated over the EHV regime for 
the three lines of our setup: CO(2--1), SiO(5--4), and 
SO(J$_{\mathrm N}$=6$_5$--5$_4$). 
The EHV regime is defined as
the 20~km~s$^{-1}$-wide velocity range
that is centered on $|V_0\pm 37.5\mathrm{~km~s}^{-1}|$, where $V_0$ is the LSR 
velocity of the ambient cloud (6.7~km~s$^{-1}$, T04), and
the plus and minus signs correspond to the red and blue shifted parts of
the EHV regime, respectively.

As can be seen in the top panels of Fig.~\ref{fig3}, CO is detected in both target 
fields with high signal-to-noise ratios. In each target, most
of the CO emission lies inside an approximately elliptical region
that we indicated in the figure with dashed lines, and which
is further discussed below.
The northern field presents additional CO emission to the southwest, outside the
ellipse, and close to the edge of the ALMA field of view. 
This extra emission seems to
correspond to the nearby B5 EHV emission peak, which according to the estimate
by SG09, lies $15\farcs5$ SW of B6 (Fig.~\ref{fig1}). Since we have not corrected the
maps in Fig.~\ref{fig3} for the attenuation
of the primary beam of the antennas (to avoid amplifying noise near the edges), the 
emission from the B5 peak is significantly weakened and distorted in the image.

In contrast with CO(2--1), the SiO(5--4) emission
is barely detected inside the elliptical regions 
(Fig.~\ref{fig3}, middle panels).
In the northern field, the off-center
B5 peak is clearly detected, and its emission overshadows the emission
from the target B6 peak. A brighter B5 emission peak is expected from the data of SG09, who mapped SiO(2--1) together
with CO(2--1), and found that the SiO emission from both B6 and R6 is much weaker
than the emission from the inner jet peaks. Although the 
maps of Fig.~\ref{fig3} suggest that there is negligible 
SiO(5--4) emission in the elliptical regions, this is somewhat exaggerated by
integrating the signal over the 20~km~s$^{-1}$-wide
EHV regime. As we see in the next section, 
the EHV regime has a large-scale velocity gradient and the intrinsic
velocity width of the emission at any position is significantly lower
than the full width of the EHV range.
As a result, integrating
the emission over the full 20~km~s$^{-1}$-wide EHV regime dilutes the signal
and decreases the sensitivity to weak features. 
Narrower channel maps show some SiO
emission inside the elliptical regions of both fields, 
but the behavior of this emission seems so similar to that of the much brighter CO 
(presented in the next section)
that its analysis 
adds very little to the study of our target fields
(see Appendix~\ref{sio_6v} for channel maps of the SiO emission).

The last transition in our setup, SO(J$_{\mathrm N}$=6$_5$--5$_4$),
was not detected inside the elliptical regions
of either ALMA field, even when using narrow channel maps
(Fig.~\ref{fig3}, bottom panels).
The B5 peak was marginally detected 
inside a narrow velocity range, but its  signal is strongly 
attenuated by the ALMA primary beam.
Because of its weak signal, the SO emission is not further discussed here.

\subsection{Velocity structure: Evidence of expanding gas disks}

The spectra in Fig.~\ref{fig2} show that
the red and blue EHV components are symmetrically offset
in velocity with respect to the ambient cloud (at V$_{\mathrm{LSR}} = 6.7$~km~s$^{-1}$).
We estimated the value of this offset using a 
a Gaussian fit to the spectra, deriving a velocity of 37.5~km~s$^{-1}$
for both the blue and red EHV components.
This offset velocity likely represents the mean radial velocity of the IRAS~04166 jet, and as
a first step in our analysis, we corrected
the EHV emission in each ALMA field by this value.
In this way, our study concentrates on the
internal motions of the EHV gas
and measures them in the local rest frame of the jet.

Having set the mean gas velocity to zero in each field, we now divide the 
20~km~s$^{-1}$-wide EHV regime into multiple velocity channels.
Fig.~\ref{fig4} shows the result for the case of six equally spaced
channels of 3.33~km~s$^{-1}$ width, although other width choices
produce similar maps. 
As the figure shows, the two ALMA fields present very similar velocity patterns.
The emission at the lowest (bluest) velocities lies toward the
southwest boundary of the dashed ellipse, and as the velocity increases, the
emission moves gradually from SW to NE sweeping the full elliptical region. 
Finally, the reddest EHV emission lies near the
NE boundary of the ellipse and appears approximately
mirror-symmetric with respect to the extremely blue emission.
This systematic SW-to-NE shift of the emission with increasing
velocity is not only seen in the emission inside the two ellipses.
It can also be seen 
in the off-center B5 peak of the northern field, although the
pattern here is less prominent
than in B6 and R6 owing to the strong effect of the primary-beam attenuation.

The systematic shift of the emission as a function of velocity
matches the sense of the velocity gradients found by SG09 and W14 in their PV diagrams along the jet axis.
These PV gradients show that in each EHV peak, 
the absolute value of the radial velocity decreases with distance to the source.
As Fig.~\ref{fig4} shows, this is the same sense of the velocity gradient
found in the channel maps:
the reddest gas of the red (southern) field is found at positions
closer to the central source (toward the NE), and the bluest gas in the blue (northern)
field is also found toward positions closer to the source (in this case toward the SW).
The gas motions seen in Fig.~\ref{fig4} therefore represent the
velocity structure underlying the gradients previously seen in the PV diagrams.

The velocity gradients in the PV diagrams were interpreted by SG09 and W14 as arising
from the sideways ejection of gas in a series of internal jet shocks.
This interpretation was based on a comparison of the observed PV diagrams with
synthetic diagrams predicted by numerical models of pulsating jets, such as those
of \citet{sto93} (see also \citealt{voe99} and \citealt{mor16}).
The high level of detail provided by the new ALMA images
allows us now to further investigate the lateral ejection interpretation by studying the
detailed 2D distribution of the gas velocity field. To do this, we first characterize the
elliptical regions that bound the EHV emission in each ALMA field.
We use for this the maps of extreme red- and blueshifted velocity in Fig.~\ref{fig4} because they
seem to best constrain the curvature of the ellipses.
For the northern field, we estimate an ellipse size of about $20''\times 12''$
(major times minor axis), while for the southern field we estimate a slightly larger
size of $22''\times 14''$. 

\begin{figure}
\centering
\resizebox{8cm}{!}{\includegraphics{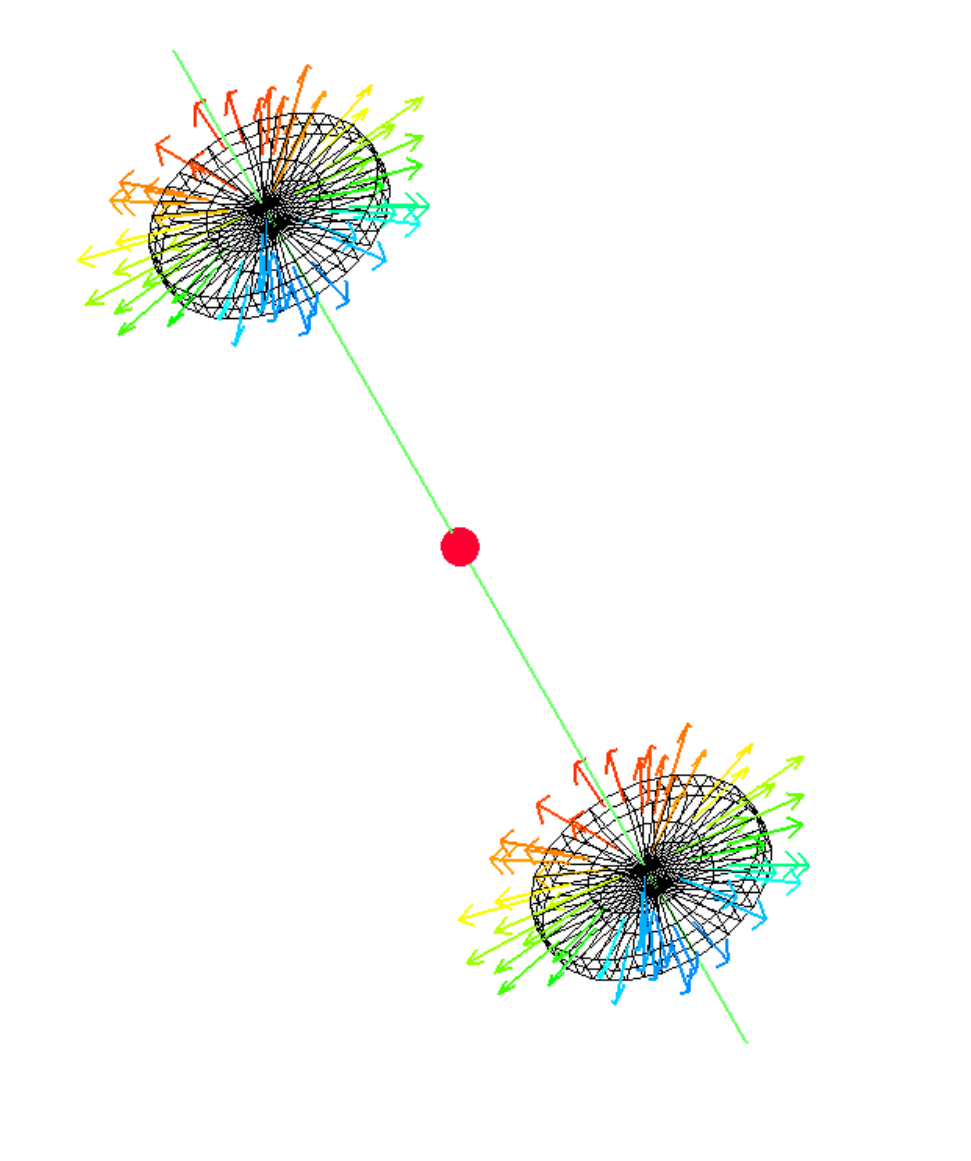}}
\caption{Schematic geometry of the emitting gas.
The filled red circle indicates the position of IRAS~04166, and the green line
is the direction of propagation of the jet. Each emitting region
is represented by a tilted disk of gas that expands perpendicular
to the jet axis, and whose velocity vectors have been color coded according to their
Doppler shift. This schematic geometry simultaneously reproduces the
elliptical shape of the emission and the systematic velocity pattern.
(To improve visibility, the size of the disks has been exaggerated
in proportion to their distance from the central source.)
}
\label{fig_disk}
\end{figure}

The final parameter needed to characterize the elliptical regions 
is the position angle (PA). For both ALMA
fields, we estimate a value of about $-60\degr$, measured clockwise from north 
according to the standard convention. This PA value indicates that
the minor axis of each ellipse is approximately parallel to
the direction of the IRAS~04166 jet, which was determined as 
$30\fdg4 \pm 0\fdg2$ by SG09 from a fit to the emission centroids of all the EHV peaks in the outflow. Such a coincidence between the
ellipse minor axis and the direction of the jet strongly
suggests that the elliptical shape of the regions is the result
of a projection effect and that the true distribution of the
gas is in circular disks that are perpendicular to the jet axis. 
This geometry is illustrated in Fig.~\ref{fig_disk},
where the gas in the two ALMA fields has been represented by 
two disks of expanding gas. As can be seen, the circular disks are
foreshortened into ellipses that have their minor axes parallel to
the direction of the jet. In addition, the expansion velocity field
of the gas creates a pattern of velocities in which 
the blueshifted material occupies
the SW section of the disk and the redshifted material occupies the NE 
section.
This pattern matches the behavior of the velocity field
seen in the channel maps of Fig.~\ref{fig4}.

If the elliptical shape of the emission in each
ALMA field results from the foreshortening of a
circular gas disk, the aspect ratio of the ellipse is an indicator
of the inclination angle of the disk with respect to the
plane of the sky. From this angle, we can estimate the
inclination angle of the jet. As we have seen, the two ALMA fields are
fitted with ellipses of very similar aspect ratio, 6/10 and
7/11, and these ratios correspond to jet inclination angles of 
$53\fdg1$ and $50\fdg5$. These values are very similar, especially
considering our simple fitting method, so we average them
as $52\degr$, and assume a probable uncertainty of 
less than 10\%. With this angle we can deproject the
radial jet velocity derived from our Gaussian
fit to the EHV emission (37.5~km~s$^{-1}$) and obtain a
true jet velocity of about 61~km~s$^{-1}$, again with an
uncertainty of about 10\%.

\begin{figure}
\centering
\resizebox{\hsize}{!}{\includegraphics{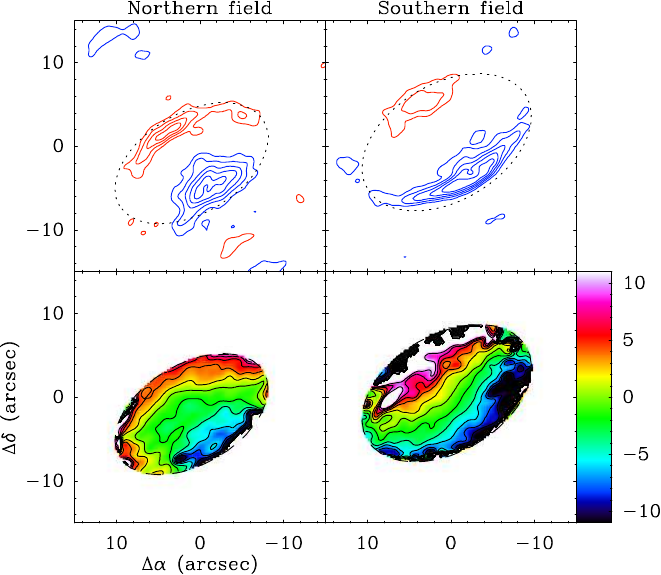}}
\caption{Evidence of curvature in the CO(2--1) emitting regions.
{\em Top panels:} comparison between the CO emission in the 
reddest and bluest 5~km~s$^{-1}$ intervals of the EHV range
for each ALMA field.
First contour and contour interval are 0.5~K~km~s$^{-1}$.
{\em Bottom panels:} first moment maps of the CO(2--1) emission.
As in Fig.~\ref{fig4}, the 
velocity scale has been centered on zero by correcting for the
source and jet velocities.
Data outside the elliptical regions have been masked out as a result of their
low signal-to-noise ratio.
The wedge to the right shows the scale in km~s$^{-1}$}
\label{fig5}
\end{figure}

\subsection{Asymmetry at high velocities: Evidence of curvature}

So far we have stressed the similarities between
the emission in the northern and southern ALMA
fields in terms of velocity offset from the ambient cloud, elliptical
shape, and velocity pattern.
These similarities are indeed the dominant feature of the emission. 
A close inspection of the data, however, also reveals slight
but systematic differences between the emission from the two ALMA fields.
One indication of these differences comes from comparing
the emission at the most extreme velocities;
 figure~\ref{fig5} shows 
together the emission in the reddest and bluest 5~km~s$^{-1}$ of the
EHV range (top panels). As expected, in both fields the blue emission lies
toward the SW 
and the red emission lies toward the NE. This blue-red symmetry, however,
is not perfect. In the northern field, the 
redshifted emission is significantly narrower and more curved than the 
blueshifted emission.
In the southern field, the asymmetry is reversed:
the blueshifted emission is
the one that is narrow and curved, while the 
redshifted emission is wider and rounder. 

The asymmetry seen at extreme velocities 
is part of a global pattern that also occurs at intermediate velocities.
This is illustrated in the bottom panels of Fig.~\ref{fig5} using
maps of the first moment (velocity centroid) of the CO(2--1) emission. The maps show the expected
layered distribution that has the blue
emission toward the SW and the red emission toward the NE. 
In addition, the maps show that in each field, the
iso-velocity contours are slightly curved and the sense
of curvature is opposite in the two fields. In
the northern field, the contours curve slightly away from the NE
(red edge), while in the southern field, the contours curve away from the SW (blue edge).

The curvature of the first-moment maps in Fig.~\ref{fig5} implies that the 
expanding gas disks responsible for the emission in each ALMA field
are not perfectly flat. They must be slightly curved, or at
least their velocity field must present a small degree of curvature. 
To reproduce the first-moment maps, the curvature must be such 
that the edge of each disk curves back toward the central 
IRAS source. This sense of curvature again matches the prediction from numerical models
of lateral expansion of gas in an internal jet shock (e.g., \citealt{har87}), which show that
the material ejected from the jet curves back because it is slowed
down by its interaction 
with lower velocity gas outside the jet beam. Numerous observations of
optical and IR jets show this type of curvature, both at the leading
end of the jet and at the internal shocks, likely caused by jet pulsations 
(e.g., \citealt{dev97,zin98,rei02}).

To summarize, the new ALMA observations support the previous interpretation
of the EHV emission as arising from gas that has been laterally ejected 
in a series of jet shocks (SG09 and W14). In addition, the 
new data provide a 
detailed picture of the distribution of gas both in space and velocity, 
thanks to the significant increase in sensitivity afforded by ALMA.
In the next section we present a geometrical model aimed at taking
advantage of this new sensitivity to further constrain the properties
and kinematics of the EHV gas.

\section{A geometrical model of the EHV gas}
\label{sec_model}

The goal of our model is to reproduce the behavior of the CO
emission in the two ALMA fields by assuming that it arises from
laterally-expanding material in jet bow shocks.
Since the emission from the two ALMA fields looks similar 
in the velocity maps (Fig.~\ref{fig4}) and moment maps
(Fig.~\ref{fig5}), we assume that the two emitting regions 
have the same internal structure, and that 
they only differ in their relative orientation
with respect to the central source.
By doing this, our model does not attempt to fit the individual details of each ALMA field, but
fits the common features of their emission.

\begin{figure}
\centering
\resizebox{7cm}{!}{\includegraphics{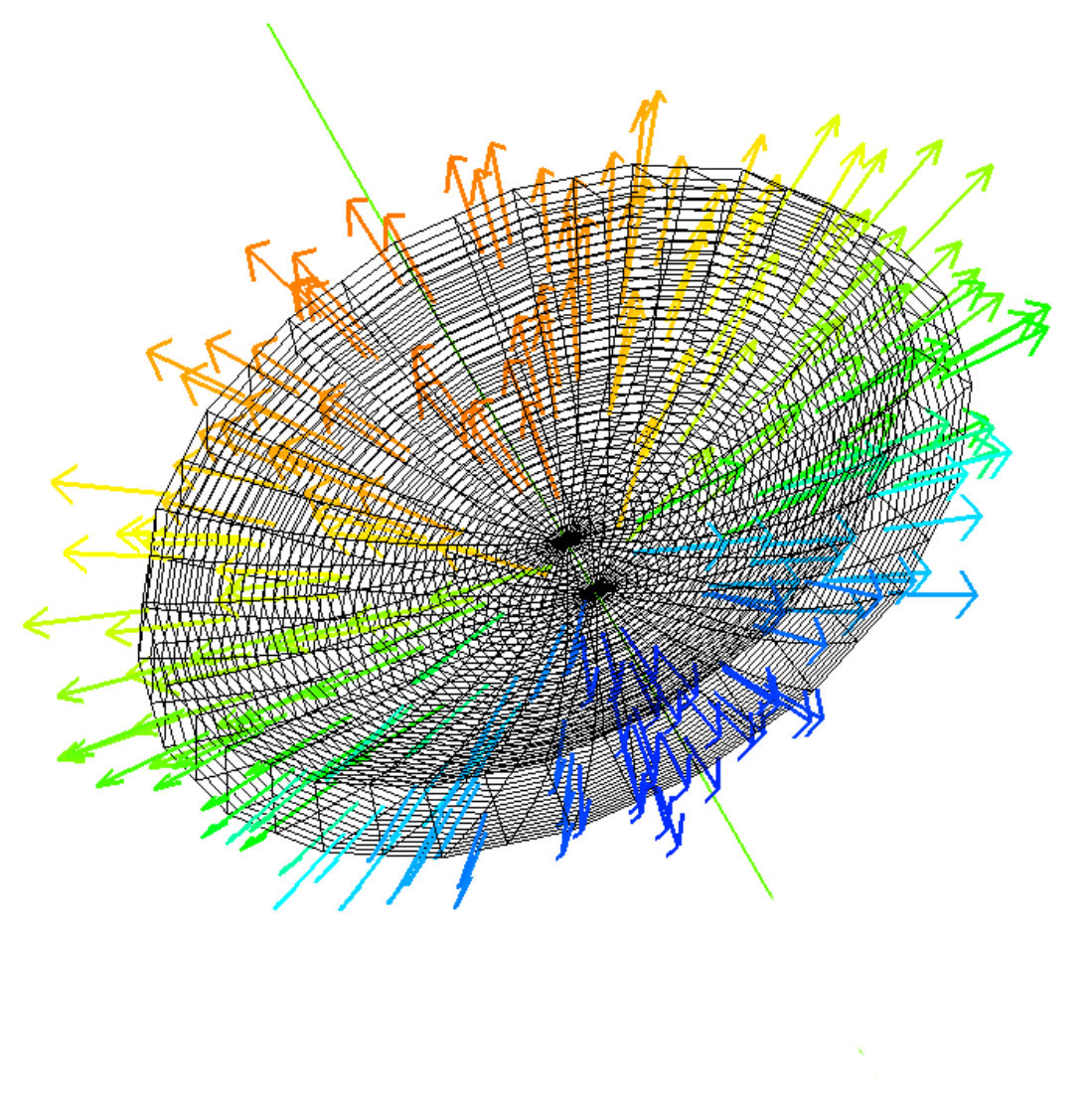}}
\caption{Schematic view of the shell used to model the emission in the
southern ALMA field (the northern model is just the reflection image with
respect to central source). As in Fig.~\ref{fig_disk}, the 
green line represents the jet axis and the velocity vectors have
been color coded according to their Doppler shift.
}
\label{fig_bow}
\end{figure}

We start our modeling by assuming that the emitting gas is distributed in a shell
of parabolic walls, and that the gas expands outward moving parallel to the walls.
This geometry is illustrated in Fig.~\ref{fig_bow} for the
southern field and 
is motivated by the analytic
shock models of \citet{har87},
who found that bow shocks tend to be parabolic
in the vicinity of the jet axis.
To reproduce our observations, the curvature of the parabola must be very mild 
(described as $z = 0.02 r^2$), and the 
shell must be truncated when its outer
cylindrical radius $r$ reaches a size of $11''$ to match the observed emission size.
Our model also assumes
that the shell has a thickness of $1\farcs 5$. This choice is not
very critical, since the emission is optically thin and depends only
on the column density. Any other choice of the shell thickness 
can be compensated with a change in the gas density.

We also assume that the gas is isothermal. 
The single dish observations of
CO(1--0) and CO(2--1) by T04 indicate 
gas temperatures  between 7 and 20~K in the different EHV peaks, with a 
tendency for the colder peaks to be found further away from the IRAS source. 
Following this trend, 
we assume that the
gas in the shells has a constant temperature of 10~K. 
We also assume a standard CO abundance of $8.5 \times 10^{-5}$ \citep{fre82}.

Once we assume the physical conditions of the gas, we
calculate the expected CO emission using
a modified version of the ray-tracing program presented in \citet{taf97}. 
This program integrates the equation of radiative transfer over a grid of positions
assuming LTE conditions and taking the gas velocity field into account.
 We explored a variety of velocity and density laws to fit the data. 
To fit the velocity maps of Fig.~\ref{fig4}, which show that the emission 
gradually sweeps the
elliptical regions and forms a series of narrow bands approximately
parallel to the major axis, 
a constant velocity model does a poor job (Appendix~\ref{compare}), and we need to use a velocity field that
increases in magnitude with distance from shell center. 
After some trial and error, we found
a best fit using a linear velocity field where the gas expansion 
increases gradually from 0 to 13~km~s$^{-1}$ at the shell edge.

The gas density law is less well constrained than the velocity, 
since it depends on our assumption
of the shell thickness, and also because the maps of integrated emission do
not show a systematic pattern but significant
differences between the two ALMA fields (Fig.~\ref{fig3}). 
To reproduce the common gradual decrease of intensity toward the emission 
edges, we used an inverse square-root density law in the outer half of the shells,
and to avoid creating a deep hole at the center, we assumed that the inner
half of the shell has a constant density of a $0.8\times 10^4$~cm$^{-3}$.
While this choice may not be unique, it seems to capture the general distribution
of mass inside the shells.

\begin{figure*}
\centering
\resizebox{\hsize}{!}{\includegraphics{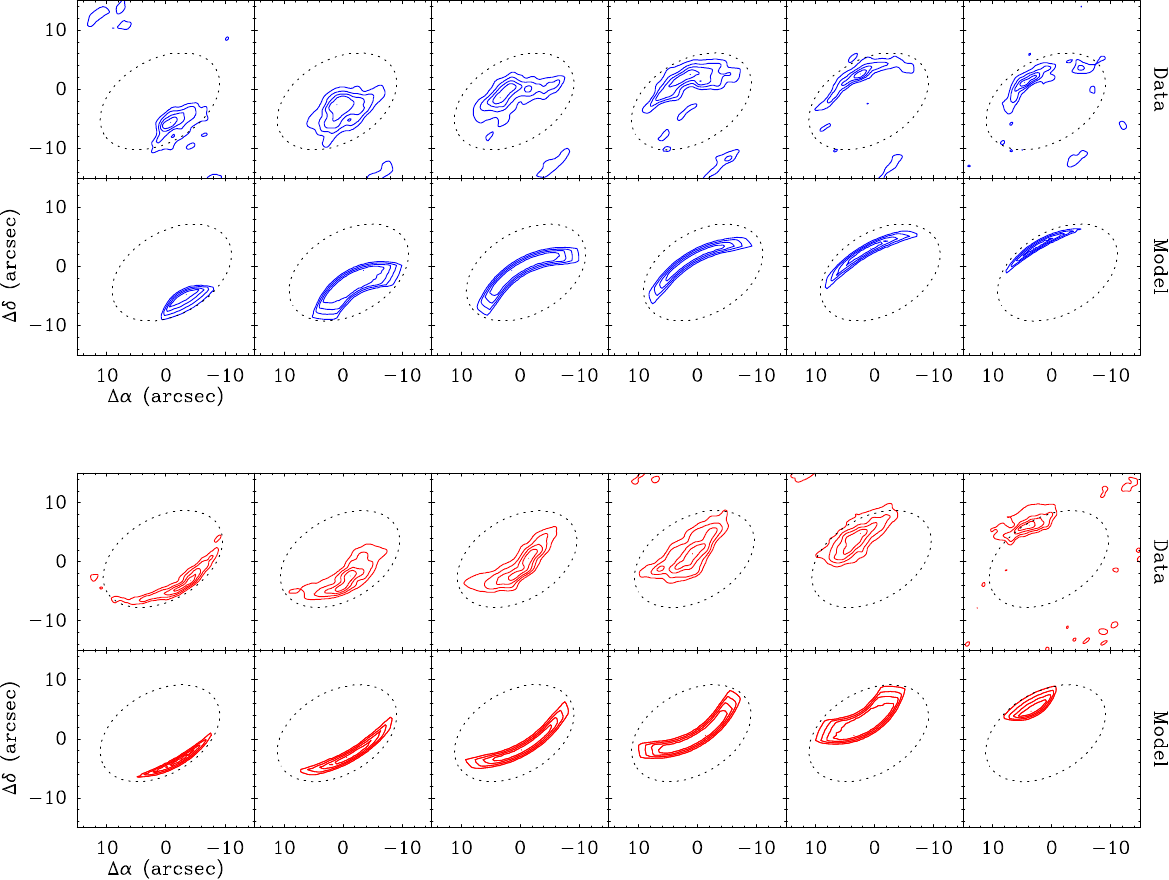}}
\caption{Comparison between observed velocity maps
and results from a geometrical model 
for both the northern (top and blue) and southern (bottom and red) ALMA fields.
Both fields have been modeled assuming the same physical conditions but opposite orientation, 
as expected for two bow shocks 
moving away from the central IRAS source. 
Panel velocities, dashed ellipses, and contour levels as in Fig.~\ref{fig4}.
}
\label{fig6}
\end{figure*}

\begin{figure*}
\centering
   \begin{minipage}[b]{0.4\textwidth}
      \includegraphics[width=0.95\textwidth]{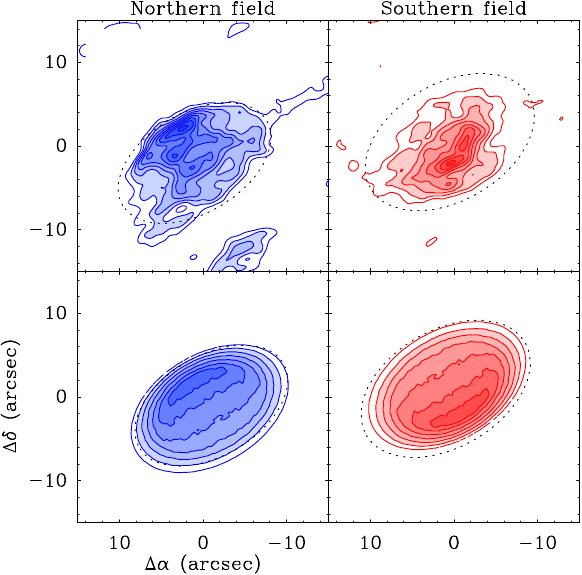}
   \end{minipage}
   \begin{minipage}[b]{0.4\textwidth}
        \includegraphics[width=0.943\textwidth]{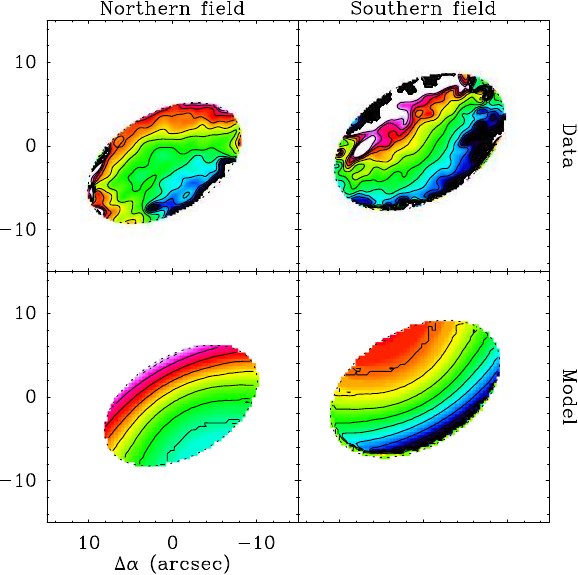}
   \end{minipage}
\caption{Comparison between maps of integrated intensity (left) and iso-velocity (right)
from ALMA observations and the geometrical model presented in the text. For each comparison, the
top panels represent the ALMA observations and the bottom panels
represent the model.
Each comparison uses the same velocity interval and contour levels
as in Fig.~\ref{fig3} (integrated maps) and Fig.~\ref{fig6}
(iso-velocity maps).
}
\label{fig_data_model_app}
\end{figure*}

Figs.~\ref{fig6} and ~\ref{fig_data_model_app} compare the results of our geometrical
model with the velocity maps, integrated maps, and first-moment maps
of the two ALMA fields.
Despite its simplicity,
the model reproduces well the main features of the
CO emission, especially its elliptical distribution, 
gradual displacement from SW to NE
as a function of velocity, and
opposite sense of curvature in the first-moment maps.
These characteristics arise from the disk-like distribution of the
emitting gas and from its expansion velocity field,
and constitute the best-constrained properties of the EHV gas.
Their success in modeling the data provide a
further confirmation that the EHV emission 
arises from the internal bow shocks of a pulsating jet.

While the geometrical model fits the data well, the agreement is not perfect.
Fig.~\ref{fig6} shows that the
model emission is significantly more concentrated than observed with ALMA, an
effect that is especially noticeable in the maps of intermediate velocities.
This discrepancy probably results from our simplified velocity field,
which assumes that the gas moves along the shell with no shear and
no internal velocity dispersion.
This is of course an extreme approximation, since 
the gas must have additional
motions resulting from the dissipation of the 
supersonic jet velocity in the 
internal shock. Such complex turbulent motions
are indeed seen in numerical simulations of jet shocks, and
arise from a number of instabilities \citep{blo90,sto93}.
Adding these motions,
likely broadens the emission in the channel maps by 
increasing the number of shell points that contribute to any given map.

Despite the above limitation, it seems clear from the model 
that the kinematics of the CO-emitting gas is dominated by the motions of 
lateral expansion along the shells. We can therefore use the model fit
to constrain the main properties of these motions. 
An important constraint from the model
is the need for the gas velocity to increase in magnitude 
with distance from the jet axis (Appendix~\ref{compare}).
This linear velocity gradient 
could arise from different physical processes.
One possibility is that the strength of the jet shock
has decreased systematically with time, and that the linear 
velocity gradient reflects this gradual shock weakening. Both
numerical and analytic models of pulsating jets with
sinusoidal velocity oscillations 
show that the strength of the shock is highest at early times,
since the original sinusoidal velocity oscillation quickly
steepens into a saw-toothed profile that makes the fastest jet gas
encounter the slowest gas from the previous cycle,
producing the strongest possible shock \citep{kof92,sto93,smi97,sut97}. 
As time progresses, the relative velocity between the incoming and 
trailing parts of the jet decreases, and so does
the strength of the shock. This decrease is approximately
linear, as shown by the analytic model of \citet{kof92} (see their
Fig.~2).

An alternative interpretation of the linear velocity gradient is that it
is the result of
a single shock event that generated gas moving at multiple
expansion velocities. In this case, the expanding material would naturally sort itself
spatially because each gas parcel would move away from the jet axis proportionally
to its expansion
velocity. To test the consistency of this interpretation,
we estimate the
time since the shock event took place by dividing the maximum 
displacement of the gas from the jet axis ($\approx 11''$) by the largest
estimated gas velocity ($\approx$ 13~km~s$^{-1}$). The result is approximately 550~yr,
which is lower than the estimated travel time of the jet gas from the protostar
($\approx 900$~yr). This means that an explosive event could have occurred
during the flight time of the jet material and, therefore, that the
interpretation as a single internal shock event 
is consistent with the rest of jet properties.

While the origin of the linear velocity gradient is unclear, the pattern 
occurs in both ALMA fields and is hinted 
in the off-center B5 peak of the northern field, so it seems to 
represent a robust characteristic of the EHV emission.
Further ALMA observations of the inner EHV peaks along the
jet, together with more detailed numerical models of the internal
shocks in pulsating jets, are clearly needed to understand this effect.

\section{Implications for outflow evolution}

The lateral ejection of material in the IRAS~04166 jet has potential consequences 
for the evolution of both the outflow and its surrounding dense
core. To evaluate these consequences, we now use the physical parameters of
the ejected gas that we have estimated with the geometrical model. As a first step, 
we integrate the model density profile of the shell to
estimate the amount of mass in each EHV peak.
For simplicity, we ignore the slight 
curvature of the shell and we treat the emitting region as a flat disk of $1\farcs 5$
thickness, $11''$ radius, and the density profile described in the 
previous section. In this way, we derive a shell mass of
$8.6 \times 10^{-5}$~M$_\sun$, which applies to each ALMA field because by construction we have fitted them using the same model.
This value is in good agreement with the
mass estimates for the B6 and R6 peaks made by SG09 by direct integration
of the CO(2--1) intensity, assuming optically thin conditions, which were
$8.8 \times 10^{-5}$ and $5.9 \times 10^{-5}$~M$_\sun$, respectively.
The agreement between the two estimates is not surprising, since 
they both make use of
CO(2--1) observations of consistent intensity and assume similar
excitation conditions. Still, it provides a further test 
that our simple geometrical reproduces the
main properties of the observations.

Using the mass estimate, we calculate the
forward momentum in the EHV shell. This is an important outflow parameter,
since the momentum is a conserved quantity in contrast with 
the kinetic energy, which is partly radiated away in shocks. 
We again assume that the shell is a flat disk, ignoring the small
curvature effect. We do this not only because the
curvature is very small, but also because 
bow-shock models show that the curvature is
caused by the transfer of momentum from the shell
to the surrounding gas in the jet ``shroud'' \citep{mas93};  our estimate refers to the
momentum originally available in the EHV gas. Since we have estimated that the shell 
moves in the jet direction with a bulk deprojected velocity
of 61~km~s$^{-1}$ (Sect.~3.2), we estimate that the
linear momentum of each EHV shell is $5.2 \times 10^{-3}$~M$_\sun$~km~s$^{-1}$.

More unique to our analysis is the estimate of the momentum 
perpendicular to the jet direction. To calculate this parameter, 
we again use the shell model
and integrate the product of the density times the radial velocity,
again assuming that the gas lies in a flat disk. The result indicates that the
sideways momentum of the EHV shell is
$7.1 \times 10^{-4}$~M$_\sun$~km~s$^{-1}$.
Dividing the estimate by the forward momentum, we find a small ratio of 0.14,
which is consistent with the high degree of
collimation found for the EHV component (T04, SG09, W14).

While significantly smaller than the forward momentum, the lateral momentum 
of the EHV gas is not negligible. To estimate its effect on the
rest of the outflow and surrounding core, we need to evaluate
the accumulated action of all EHV ejections over the outflow lifetime. This is highly uncertain because we ignore
how many EHV ejections have taken place, 
or even for how long the IRAS~04166 jet has been active. 
Since the ejections we have studied correspond to B6 and R6 in the notation
of SG09, and these authors estimated that the kinematic age of these ejections 
is about 900~yr,
we estimate that the typical time between EHV ejections is 150~yr. 
To estimate the outflow lifetime, we use the kinematic age of
3,000 yr derived by T04 from the position and velocity of the outermost EHV peaks (not covered by SG09). 
This age is probably a lower limit because the IRAS~04166 outflow
seems to be significantly larger than the region covered by T04,
as shown by the larger scale map of \citet{nar12},
and also because kinematic distances tend to underestimate
the true age of outflows (e.g., \citealt{par91}). Still, using the above 
numbers, we estimate that there have been a total of 20 EHV ejections (in each direction)
over the history of the IRAS~04166 outflow. If all these ejections have had the
same lateral momentum as those observed with ALMA, we estimate a total injection
of lateral momentum  by the jet of about $1.4 \times 10^{-2}$~M$_\sun$~km~s$^{-1}$
in each outflow direction.

To evaluate the impact of this momentum deposition on the surrounding core,
we first estimate the amount of core material that has been evacuated by the outflow.
According to SG09 and W14, the IRAS~04166 low-velocity outflow moves along the
walls of a cavity
whose full opening angle varies from 40-50$\degr$ close to the
central source to about $32\degr$ at larger distances.
Assuming now that the core is spherical and that the outflow cavity is a spherical sector
with a full opening angle of $40\degr$ (intermediate between the inner and
outer estimates of SG09 and W14), we estimate that each
outflow cavity corresponds to 3\% of the full core volume.
Since \citet{shi00} and T04 estimated the mass of the core at between 1.0 and 1.5~M$_\sun$, 
we deduce that the mass that used to fill each outflow
cavity was approximately 0.04~M$_\sun$. 
For this mass to be affected against the recovering thermal pressure, for example, to be pushed to the sides to 
create the cavity, we estimate that the impulse necessary is on the order
of the displaced mass times the characteristic sound speed, which is
0.2~km~s$^{-1}$ for the typical core gas temperature of 10~K (T04).
This means that the estimated impulse would be $8 \times 10^{-3}$~M$_\sun$~km~s$^{-1}$
for each outflow cavity.

As can be seen, the sideways momentum imparted by the outflow over the last 3,000~yr is
almost twice that estimated as necessary 
to push aside and open the outflow cavity. The estimate,
of course, is very approximate, given the many assumptions and simplifications
in our treatment of the outflow-core interaction. Still, it indicates that lateral
ejection of material by internal jet shocks has a potential
impact on the distribution of mass in an outflow and its core. 

The idea that internal bow shocks can produce outflow cavities 
was originally proposed by \citet{rag93}, who 
argued that the gas ejected sideways in the internal shocks of a 
pulsating jet could entrain the surrounding
ambient material and give rise to the full low-velocity outflow component.
Unfortunately, the viability of this mechanism 
has not yet been systematically studied. A number 
of numerical simulations of pulsating jets have been presented in the literature,
but they tend to show that internal jet
shocks propagate sideways for only one or two jet radii before dissolving
in the turbulent wake of the jet
(e.g., \citealt{sto93,gou94,sut97,smi97,lee01}).
The EHV features of IRAS~04166 mapped with ALMA have a radius of $11''$, and are 
therefore 
significantly wider than the jet, whose radius is probably $< 1''$ according
to the high angular resolution images of W14, and in agreement with
the expected jet radius for a typical Class 0 source \citep{cab07}.
As seen in  Fig.~1 and in the larger scale maps of T04,
there is evidence of even wider 
EHV peaks in the outermost parts of the IRAS~04166 jet.

The IRAS~04166 jet is not unique in having large internal bow shocks.
Other jets with similar characteristics have been found using
optical/IR observations, as in the cases of
HH~212 \citep{zin98} and HH~34 \citep{dev97,rei02}.
These observations suggest that pulsating jets
often extend their sideways action over distances that are larger than a few jet radii
without the need of additional mechanisms (e.g., precession), 
and that numerical simulations that produce small bow shocks are either 
missing some physical process or may not have been run for long enough to 
eliminate the transient effect of the jet head passage.
A new generation of pulsating jets models is clearly needed to
clarify this important issue. Additionally, 
further observations of the IRAS~04166 outflow, and of similar outflows with a well-defined
EHV component, are urgently needed to test the generality of the results
presented here, and to further clarify the still mysterious relation
between highly collimated jets and wide-angle molecular outflows.

\section{Conclusions}

We used the ALMA interferometer to observe two fields
in the IRAS~04166 molecular jet, each one centered on one EHV peak. 
From the analysis of the
CO(2--1), SiO(5--4), and SO(J$_{\mathrm N}$=6$_5$--5$_4$)
emission, we reached the following conclusions.

In the two observed fields, the EHV emission is concentrated in 
elliptical regions of similar size, aspect ratio, and 
position angle. The geometry and orientation of these regions with respect
to the previously determined direction of the jet implies 
that the emitting gas is located in two disk-like structures that are perpendicular to the jet axis. From this geometry, we 
estimate that the IRAS~04166 jet is inclined by about $52\degr$ with respect
to the plane of the sky and that its deprojected velocity 
is 61~km~s$^{-1}$.

When corrected for the jet velocity,
the emission in both fields presents similar velocity patterns.
At the lowest velocities, the emission lies along the SW edge of
the ellipse, and as the velocity increases, the emission shifts
gradually toward the NE while sweeping the elliptical region. 
This velocity pattern is best explained as resulting from the
systematic expansion of the emitting gas away from the jet axis.

Although the velocity pattern of the two ALMA fields is very similar, there 
is a small asymmetry in the curvature of the iso-velocity contours.
This asymmetry indicates that the disk-like structures responsible
for the EHV emission are slightly curved in the opposite sense in the
two fields. The sense of curvature is that expected for bow shocks
that move away from the central IRAS source.

We used a geometrical radiative-transfer 
model to reproduce the main behavior of
the EHV emission in the two ALMA fields. 
The model indicates that the ALMA observations can be
explained as resulting from the emission by two slightly curved shells
of expanding gas. The velocity of this expansion
increases linearly away from the jet axis, which may be an indication
of each emission peak 
arising from a single explosive event, although other options 
are possible.

Our observations and modeling confirm the previous interpretation
of the EHV emission in IRAS~04166 as arising from gas ejected sideways
in a series of internal shocks inside a pulsating jet.
The new ALMA data provide quantitative information on the characteristics
of this ejected gas, and in particular, its amount of linear momentum.
While the momentum is dominated by the forward component along the
jet direction, the sideways component is not negligible.
We estimate that over the lifetime of the outflow, this sideways
component could deposit enough momentum on the surrounding cloud 
to affect the distribution of gas and even produce an outflow
cavity. Further  numerical simulations of pulsating jets and 
observations of the IRAS~04166 outflow are necessary to fully
explore this possibility.

\appendix

\section{SiO(5--4) channel maps} 
\label{sio_6v}

\begin{figure*}
\centering
\resizebox{\hsize}{!}{\includegraphics{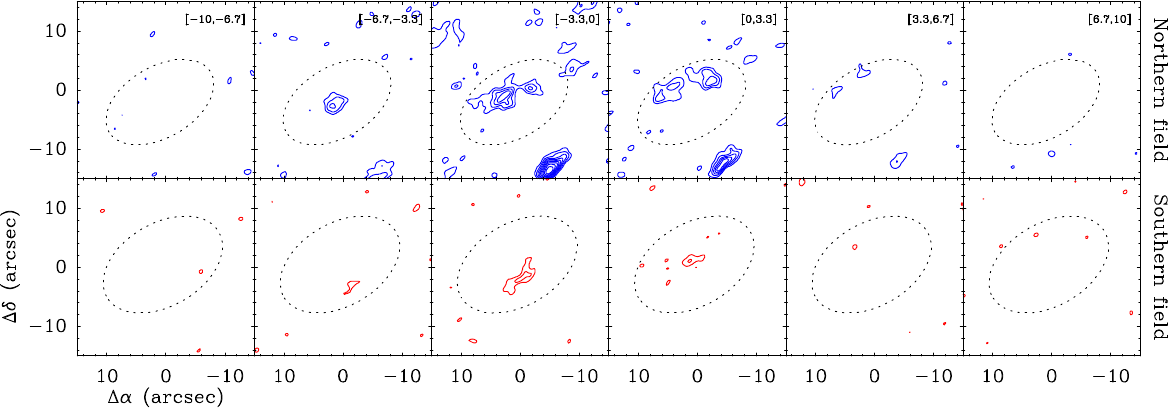}}
\caption{Channel maps of SiO(5-4) emission for the northern (top, color-coded blue)
and southern (bottom, color-coded red)
ALMA fields. Each map covers a 3.33~km~s$^{-1}$-wide velocity range indicated in the
top row. First contour and interval are at 0.15~K~km~s$^{-1}$.
Compare to the equivalent maps for CO shown in Fig.~\ref{fig4}.
}
\label{sio_vel_map}
\end{figure*}

Fig.~\ref{sio_vel_map} presents channel maps of the SiO(5--4) emission from the
two ALMA fields using the same
velocity intervals as used for CO(2--1) in Fig.~\ref{fig4}. The SiO emission
inside the ellipses
is significantly weaker than the CO emission, but the velocity pattern of the two
is similar as far as the low signal-to-noise ratio of SiO allows one to see.
The bright SiO emission to the SW in the northern field corresponds to
the B5 peak.
As already seen in CO, this emission
shows the characteristic SW-to-NE displacement with velocity seen in the
emission of the main peaks.

\section{Comparison with a constant velocity model}
\label{compare}

\begin{figure*}
\centering
\resizebox{\hsize}{!}{\includegraphics{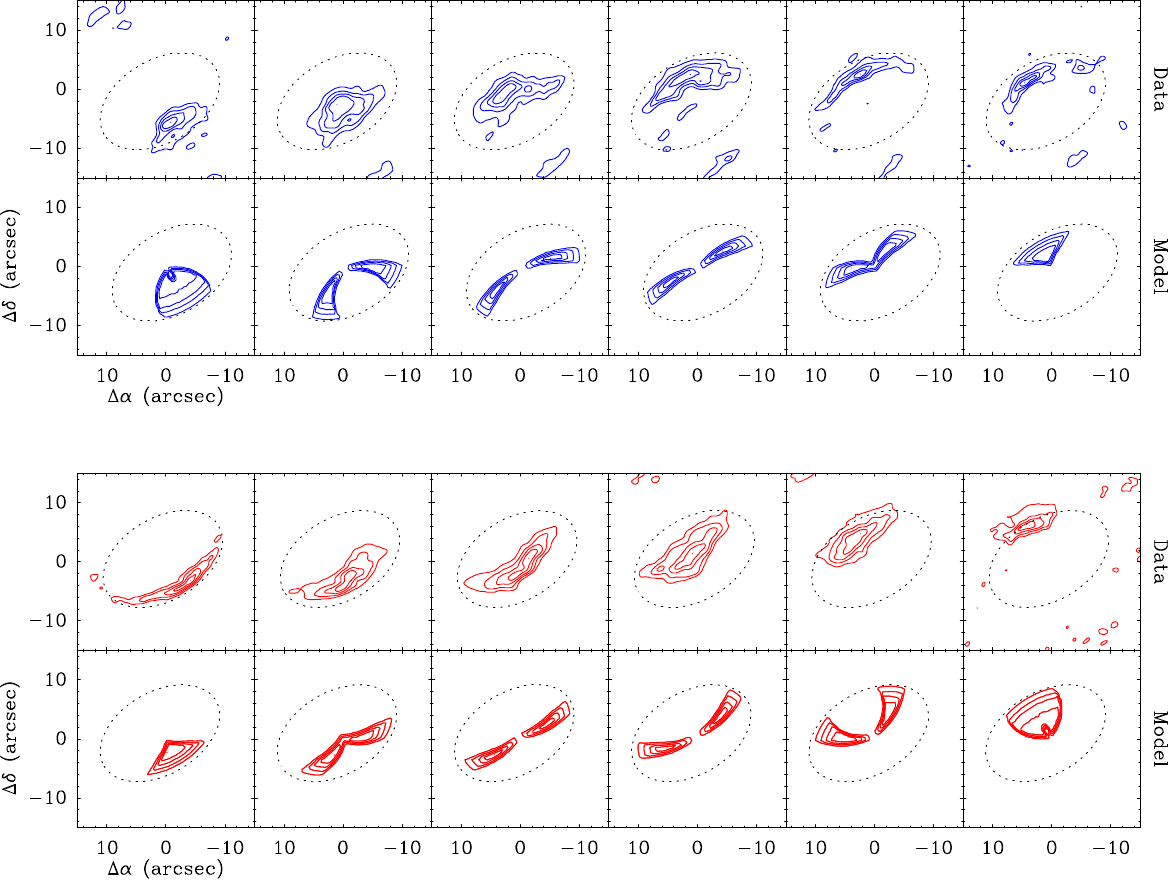}}
\caption{Comparison between observed velocity maps
and results from a geometrical model that assumes constant velocity
for both the northern (top and blue) and southern (bottom and red) ALMA fields.
All labels and contours as in Fig.~\ref{fig6}.
}
\label{constant_v_1}
\end{figure*}

\begin{figure*}
\centering
   \begin{minipage}[b]{0.4\textwidth}
      \includegraphics[width=0.95\textwidth]{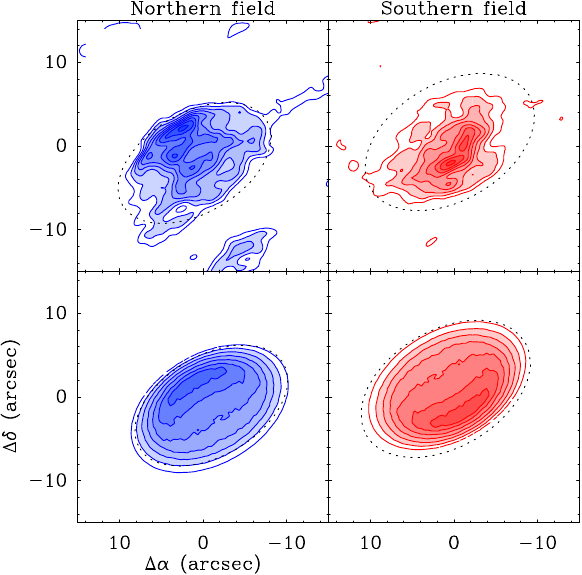}
   \end{minipage}
   \begin{minipage}[b]{0.4\textwidth}
        \includegraphics[width=0.943\textwidth]{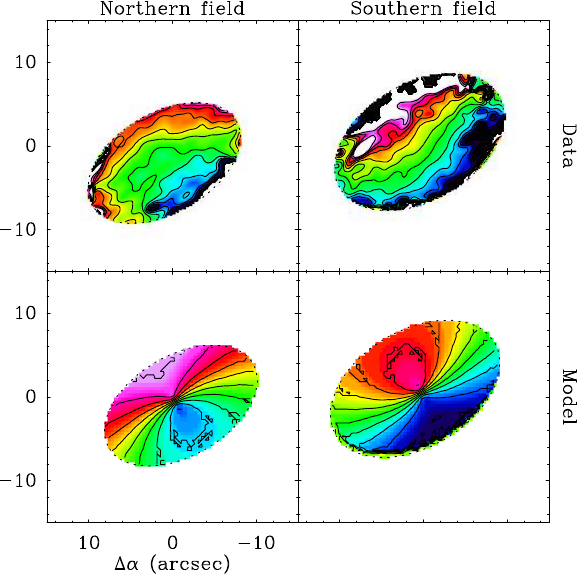}
   \end{minipage}
\caption{Comparison between maps of integrated intensity (left) and first momentum
(right) from ALMA observations and a geometrical model that assumes constant velocity. 
All labels and contours as in Fig.~\ref{fig_data_model_app}.
}
\label{constant_v_2}
\end{figure*}

To illustrate the need for a velocity gradient to fit the CO emission,
we present here the results of a model that assumes constant velocity. This model uses the same parabolic shell geometry and 
gas parameters as the best-fit model presented 
in Sect.~\ref{sec_model} with the only difference that its expansion velocity 
has a constant value of 11~km~s$^{-1}$. 
Figs.~\ref{constant_v_1} and \ref{constant_v_2}, 
compare the predictions of this constant-velocity model with the ALMA observations,
using the same maps used 
to compare the best-fit model with the data
in Figs.~\ref{fig6} and \ref{fig_data_model_app}.

As Fig.~\ref{constant_v_1} shows, the constant-velocity model predicts
channel maps where the emission is distributed in a series of wedges that
have their apex near the ellipse center and extend toward the ellipse boundary.
At intermediate velocities, two symmetric wedges appear in each map, while 
only one wedge is present at extreme velocities.

The same wedge-like distribution of the velocity field can be seen in
the first-moment map of Fig.~\ref{constant_v_2}. This figure shows
that the iso-velocity contours emerge radially from the ellipse center
and have the ellipse minor axis as a plane of symmetry.
This radial distribution of the contours 
is a direct consequence of the constant
velocity field, which generates the different values of the radial velocity
by simple projection. Since the
ALMA data are inconsistent with this pattern, we can safely exclude
a constant velocity of expansion.

In contrast with the channel and first moment maps, the integrated-intensity map
shown in the left panels of Fig.~\ref{constant_v_2} presents a reasonable fit
to the data. The map is in fact identical to that obtained with the best-fit model. This is 
expected because the emission is optically thin, so its integrated
value depends only on the gas column density, which by construction 
is the same in the two models.

\begin{acknowledgements}

We thank Miguel Santander-Garc\'{\i}a for help with the programs Shape \citep{ste11}
and Shapemol \citep{san15}, which were used for the initial modeling of the data and 
in two figures.
MT acknowledges financial support from projects FIS2012-32096 and AYA2012-32032 
of Spanish MINECO. 
DJ is supported by the National Research Council of Canada and by an NSERC Discovery Grant.
This paper makes use of the following ALMA data: ADS/JAO.ALMA\#2012.1.00304.S. 
ALMA is a partnership of ESO (representing its member states), NSF (USA), and 
NINS (Japan), 
together with NRC (Canada) and NSC and ASIAA (Taiwan) and KASI (Republic of Korea), 
in cooperation with the Republic of Chile. The Joint ALMA Observatory is operated by ESO, 
AUI/NRAO and NAOJ.
The National Radio Astronomy Observatory is a facility of the National Science Foundation 
operated under cooperative agreement by Associated Universities, Inc.

\end{acknowledgements}

%
%

\bibliographystyle{aa} 
\bibliography{29493ms.bib}

\end{CJK*}
\end{document}